\documentclass[fleqn,twoside]{article}
\usepackage{espcrc2}

\usepackage{graphicx}
\usepackage[figuresright]{rotating}

\newcommand{\AmS}{{\protect\the\textfont2
  A\kern-.1667em\lower.5ex\hbox{M}\kern-.125emS}}
\newcommand{\be}{\begin{equation}}
\newcommand{\ee}{\end{equation}}
\newcommand{\ba}{\begin{eqnarray}}
\newcommand{\ea}{\end{eqnarray}}
\newcommand{\bi}{\begin{itemize}}
\newcommand{\ei}{\end{itemize}}

\title{ Superscaling in electron- and neutrino-nucleus scattering }

\author{Maria B. Barbaro
\\
{Dipartimento di Fisica Teorica, 
Universit\`a di Torino and INFN, Sezione di Torino, Italy
}}
       
\begin{document}

\begin{abstract}
The superscaling properties of electron scattering data are used to extract
model-independent predictions for neutrino-nucleus cross sections.
\vspace{1pc}
\end{abstract}

\maketitle

\section{Introduction}

The analyses of many of the ongoing and future neutrino experiments 
require reliable predictions of the neutrino-nucleus cross sections.
Various model calculations have been performed in the quasielastic peak (QEP)
region, where the dominant process is the knock-out of a single nucleon.
However, the uncertainty due to different treatments of the nuclear
dynamics is still high when compared with the required precision.

The discrepancy between various models is even more pronounced when one moves
from the QEP to higher energy transfers, where the nucleonic resonances, 
in particular the Delta, come into play. A complete description of the 
so-called ``dip'' region between the QE and the $\Delta$ peak is in fact
still missing.

Of course any nuclear model to be applied to neutrino scattering should first
be tested against electron scattering data, where a much higher statistics 
exists than in neutrino-nucleus physics which allows to discriminate between 
different models.
It is well-known that the simple Fermi gas (FG) model, 
although accounting for the gross features of the inclusive $(e,e')$ 
differential cross sections, is inadequate for a detailed description 
of lepton-nucleus reactions. 
Specifically, the FG overestimates the ratio between the longitudinal and
transverse responses at the QEP and predicts sharp edges of the cross 
section versus the energy transfer $\omega$, in contrast with the tails 
displayed by the experimental data. 
Moreover, it predicts a linear behavior of the cross 
sections at low $\omega$ and $q$ which is not supported by the experiment. 

For this reason more sophisticated models and techniques (including RPA, 
final state interactions, short range correlations, Green's function 
methods) have been used
in the last decades to describe electron scattering reactions. 
Most of these models are based on a non-relativistic treatment
of the nuclear problem, which indeed works quite well in the low and
intermediate domain of momentum transfers up to about 500 MeV/c.
However, in the energy regime of few GeV of interest for modern neutrino 
experiments, relativistic effects cannot be ignored.

Although several efforts to account for relativistic effects 
in neutrino scattering have recently been made (see, e.g.,
Refs.~\cite{Meucci:2003cv,Amaro:2005dn,Caballero:2005sj}),
a consistent relativistic description 
of lepton-nucleus cross sections is hard to 
achieve~\cite{Amaro:2002mj}.

To overcome this difficulty, an alternative approach can be 
taken~\cite{Amaro:2004bs}:
instead of using a specific nuclear model, one can try to extract model-independent 
predictions for $\nu-A$ from $e-A$ experimental cross sections.

The method relies on the superscaling properties of the electron scattering
data: at sufficiently high momentum transfer the inclusive 
differential $(e,e')$
cross sections, divided by a suitable function which takes into account
the single nucleon content of the problem, depends only upon one kinematical 
variable, the scaling variable (this behavior is called 
scaling of first kind) and the resulting
function is roughly the same for all nuclei (scaling of second kind).
When both kinds of scaling are fulfilled the cross section is said to
superscale.

More specifically the scaling function
\be
f=\frac{d^2\sigma/d\Omega dk^\prime}{S(q,\omega)}~,
\ee
being $S$ related to the  single nucleon cross section (see 
\cite{Amaro:2004bs}), 
becomes, for large $q$, a function only of the scaling variable 
\be
\psi_{\rm QE}=\pm\sqrt{1/(2T_F)\left(q\sqrt{1+1/\tau} -\omega -1\right)}~,
\ee
where $T_F$ is the Fermi kinetic energy, $4m_N^2\tau=q^2-\omega^2$
and the $-$ ($+$) sign corresponds to energy transfers lower (higher) than the
QEP ($\psi=0$).

An extensive analysis of electron scattering data~\cite{scaling} has shown that
scaling of the first kind is fulfilled at the left of the QEP and broken at its 
right, whereas scaling of the second kind is very well satisfied at the left 
side of  the peak and not so badly violated at its right. 
As a consequence a scaling function
$f^{\rm QE}(\psi_{\rm QE})$, which embodies the scaling part of the 
nuclear effects {\it for any momentum transfer $q$
and any nucleus} (provided $q$ is high enough,
namely larger or of the order of about 400 MeV/c) 
can be extracted from the data.

The scaling violations observed at high energy transfers reside in the 
transverse response function and their main origin is the excitation of a 
$\Delta$ resonance.

The superscaling analysis has been extended to the first resonance
peak in Ref.~\cite{Amaro:2004bs}, where the contribution of the $\Delta$
has been (approximately) 
isolated in the experimental data by subtracting the quasielastic scaling 
contribution from the total experimental cross sections; the reduced
cross section has been studied as a function of a new scaling variable
\be
\psi_\Delta=\pm\sqrt{1/(2T_F)\left(q\sqrt{\rho+1/\tau} -\omega\rho -1\right)}~,
\ee
which accounts for the inelasticity through the quantity 
$\rho=1+(m_\Delta^2-m_N^2)/(4\tau m_N^2)$.
The results show that also in this region superscaling is working quite well 
and therefore a second superscaling function, $f^\Delta(\psi_\Delta)$,
can be extracted from the data to account for the nuclear dynamics.
Clearly this approach can work only at $\psi_\Delta < 0$, since at 
$\psi_\Delta > 0$ other resonances and the tail of the deep-inelastic 
scattering start contributing.

The two scaling functions extracted from $(e,e')$ data 
can then be used to predict neutrino cross sections in the 
same kinematical range.
Before showing the results, we briefly summarize the formalism for 
neutrino-nucleus scattering.

\section{Formalism}

\subsection{Charged-current reactions}

\vspace{0.2cm}
The neutrino-nucleus cross section can be written in terms of
response functions in a generalized Rosenbluth decomposition 
($q \parallel z$). For charge-changing processes 
$$\nu+A\to l+B$$
this reads~\cite{Amaro:2004bs}
\ba
&&\left[ \frac{d^2 \sigma }{d\Omega dk' }\right] _{\chi }
= 
\sigma^{(CC)}_{0}
\left\{ \widehat{V}_{CC}
R_{CC}
+2\widehat{V}_{CL}
R_{CL}
 \right. 
\nonumber
\\
&&
\left. 
+\widehat{V}_{LL} R_{LL}
+\widehat{V}_{T} R_{T}
+ 2 \chi 
\widehat{V}_{T^{\prime }}
R_{T^{\prime }}
\right\}~,
\label{sCC}
\ea
where $\Omega$ and $k^\prime$ are the scattering angle and momentum of the
outgoing lepton. 
In (\ref{sCC}) $\sigma_0$ and $\widehat V$ are kinematical factors
and $\chi=+1(-1)$ for neutrino (antineutrino) scattering.

In the Relativistic Fermi Gas (RFG) model the nucleus is viewed as a collection
of on-shell nucleons moving with energy
$E_p=\sqrt{p^2+m_N^2}$
and described by free Dirac spinors  $u({\bf p},s)$.
In spite of its simplicity the model fulfills two fundamental
requirements which are usually difficult to satisfy: Lorentz covariance
(the hadronic tensor transforms as a Lorentz tensor) and gauge invariance 
(the nuclear vector current is conserved).
Moreover it allows for analytic expressions of the response functions:
\be
R^{(i)}_{RFG}(q,\omega) 
={\cal N}\frac{m_N T_F |Q^2|}{k_F^3 q}
R^{(i)}_{s.n.}(q,\omega)
f_{RFG}(\psi)\ ,
\ee
where ${\cal N}=Z,N$ is the nucleon number, $k_F$ the Fermi momentum 
and $R^{(i)}_{s.n.}$ the ``single nucleon'' response corresponding to the
elementary reaction.
The function
\be
{f_{RFG}(\psi )}=\frac{3}{4}(1-\psi ^{2})\theta (1-\psi ^{2})
\ee
is {\em universal}:
when plotted against the scaling variable
$\psi(q,\omega; k_F)$ no dependence is left either on 
the momentum transfer $q$ or on the nuclear species via $k_F$. 
As above mentioned, we then say that the RFG superscales.
Note that the RFG response region is restricted to $-1\leq\psi\leq 1$.

The above expressions hold for both the QE and the $\Delta$ peaks,
the only differences being the single nucleon tensor (which corresponds to the
elementary reaction $\nu N\to l N$ and $\nu N\to l\Delta$, respectively) and the scaling variable,
which is different because of the different kinematics.
In particular the QE and $\Delta$ peaks correspond to $\psi=0$, which implies
$\omega^P=|Q^2|/(2 m_N)$ for QE and $\omega^P=(|Q^2|+(m_\Delta^2-m_N^2))/(2 m_N)$ for the
$\Delta$ peak.

The RFG model approximately reproduces the shape and position of the two peaks,
providing an energy shift, which accounts for the nucleon
separation energy, is introduced
in the model through the substitution $\omega\to\omega-E_{shift}$, entailing
$\psi\to\psi^\prime$. Note that $E_{shift}$, typically of the order of 20 MeV, 
and the Fermi momentum $k_F$ are the only parameters of the model.

To go beyond the RFG according to the procedure illustrated in the previous section,
we just replace the superscaling function $f_{RFG}$ with  
$f^{\rm QE}$ and $f^\Delta$, extracted directly from the experiment 
in the QE and $\Delta$ regions, respectively.

\subsection{Neutral-current reactions}

\vspace{0.2cm}
For neutral current (NC) processes 
$$\nu+A\to \nu^\prime+(A-1)+N$$
an expression similar to (\ref{sCC}) holds~\cite{NC}:  
\ba
&&\left[ \frac{d^2 \sigma }{d\Omega_N dp_N }\right] _{\chi }
= 
\sigma^{(NC)}_{0}
\left\{{V}_{L}
R_{L}
+{V}_{T}
R_{T}
 \right. 
\nonumber
\\
&&
+{V}_{TT} R_{TT}
+{V}_{TL} R_{TL}
\nonumber
\\
&&
\left. 
+2\chi \left(V_{T^\prime}R_{T^\prime}+V_{TL^\prime}R_{TL^\prime}
\right)\right\}~,
\label{sNC}
\ea
where $\Omega_N$ and $p_N$ are the scattering angle and momentum of the emitted nucleon and
now 6 response functions appear.

In this case the final state which is detected 
is the outgoing nucleon, and the neutrino kinematic variables are integrated 
over. Therefore the kinematics of the reaction is different from the one
of CC processes and of $(e,e^\prime)$: 
these are $t$-scattering type processes (the Mandelstam variable $t$ is fixed), 
whereas the neutral current reactions are $u$-scattering processes (the 
variable $u=k-p_N$ is fixed and $t$ is integrated over)~\cite{Barbaro:1996vd}.
As a consequence, it is not obvious that the superscaling procedure, based
on the analogy with inclusive electron scattering, is still valid.

This issue is discussed at length in Ref.~\cite{NC}, where it is shown
that the scaling method is based on a factorization assumption which has to be 
tested
numerically. Actually, as illustrated in \cite{NC}, 
the scaling method appears to be
applicable also to neutral current reactions, at least in the QEP region. 
The resonance production would be more involved, 
since the final state involves both a nucleon and a pion.

\section{Results}

Before showing the results for neutrino reactions we 
test the validity of the function $f^{\rm QE}(\psi_{\rm QE})+f^\Delta(\psi_\Delta)$ as a 
representation of the nuclear dynamics for electron scattering in the
quasielastic and $\Delta$ peaks.
In Fig.~1 the cross section reconstructed by multiplying the phenomenological
superscaling function by the appropriate single nucleon functions is compared with
the data for different kinematics and different nuclei. It appears that typical deviations 
are 10\% or less, thus confirming that the scaling approach offers a reliable 
description of the nuclear dynamics.
\begin{figure}[ht]
\begin{center}
  \includegraphics[scale=0.4,clip,angle=0]{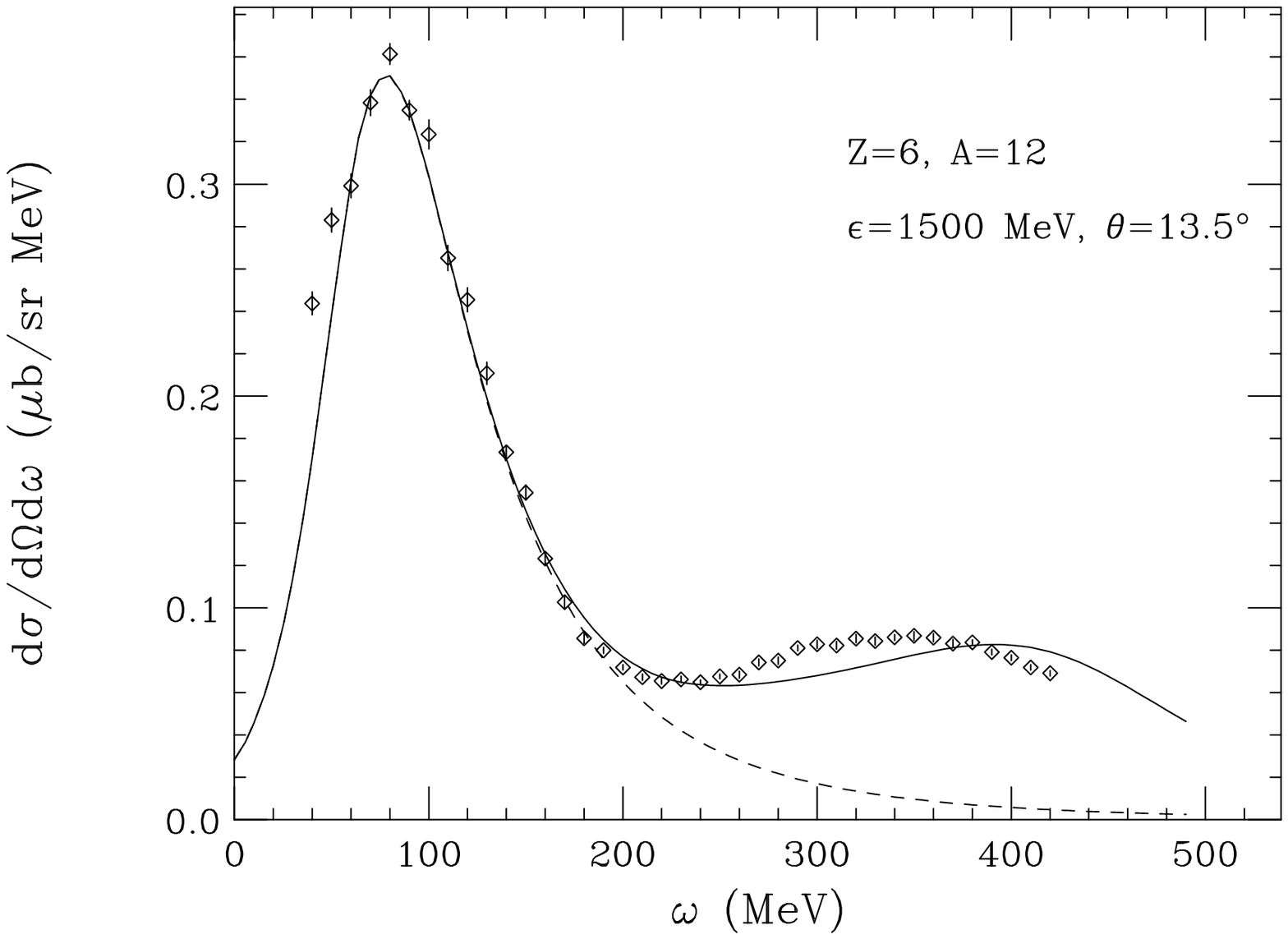}
  \includegraphics[scale=0.4,clip,angle=0]{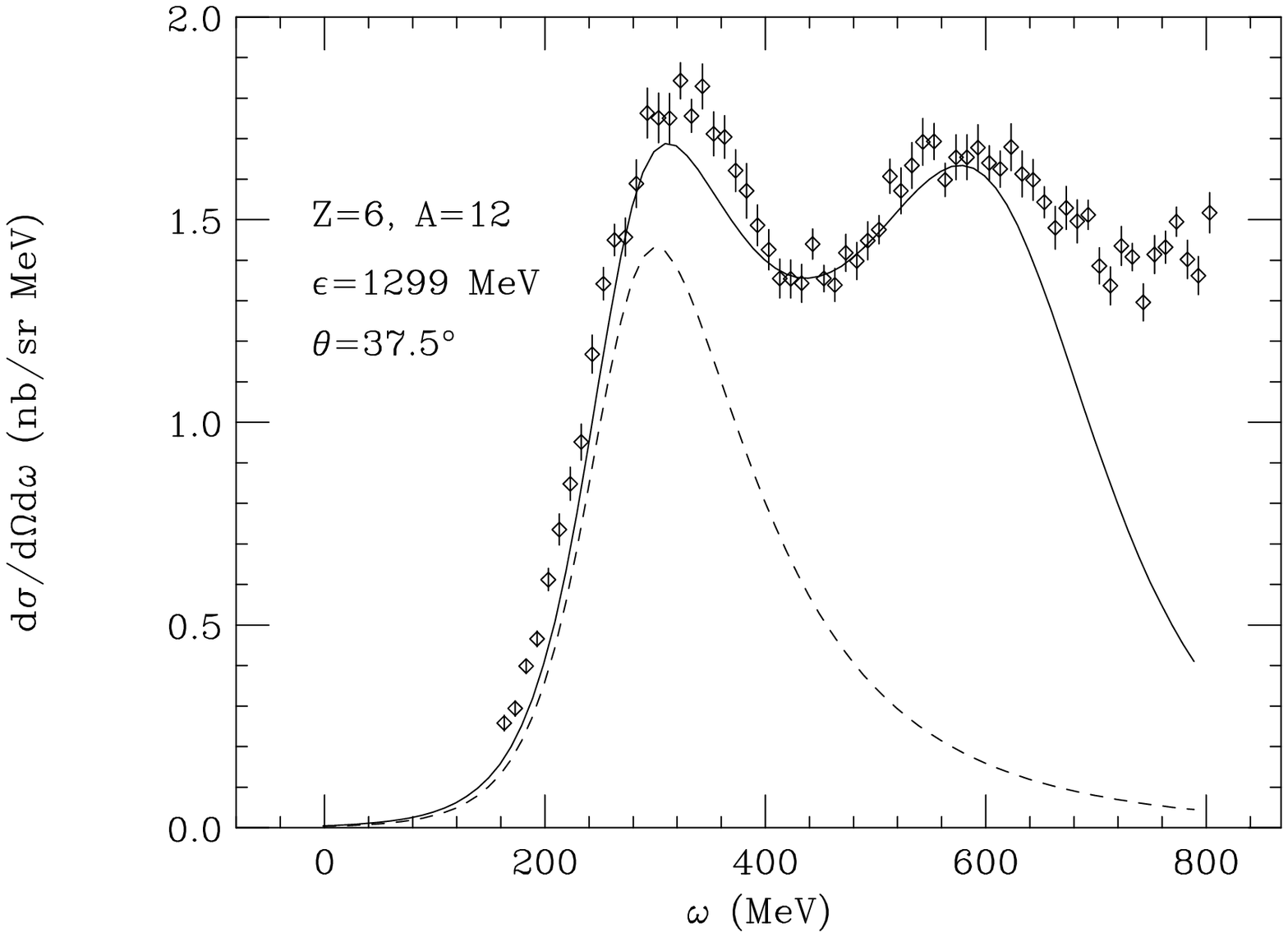}
  \includegraphics[scale=0.4,clip,angle=0]{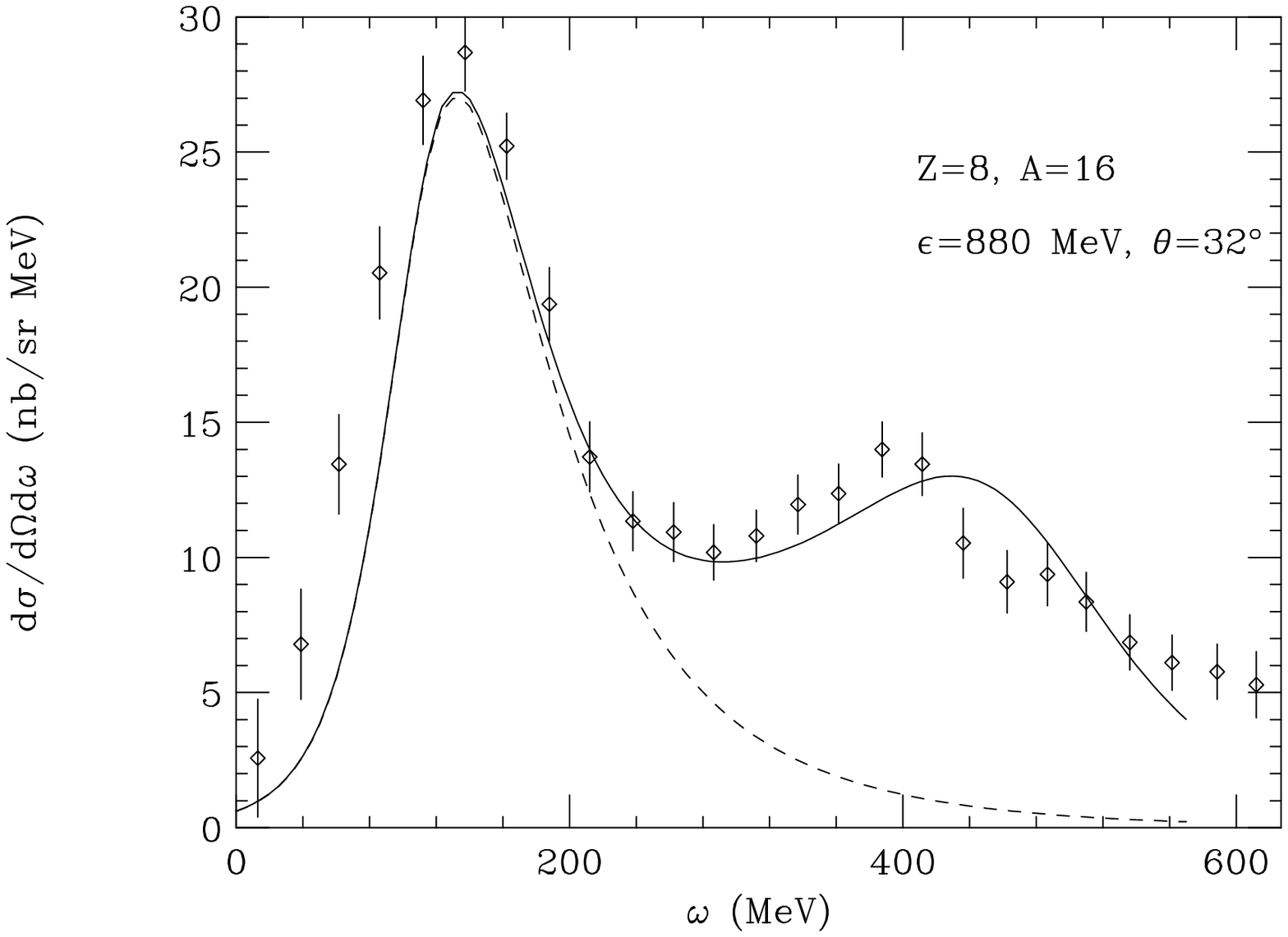}
  \parbox{6cm}{\caption[]{ Experimental $(e,e')$ cross section for $^{12}$C and
$^{16}$O at different incident
      electron energies and scattering angles. The solid curves represent
      the calculated result obtained using $f^{\rm QE}$ and
      $f^\Delta$, whereas the dashed curves are the QE contributions. (Ref.~\cite{Amaro:2004bs})
  }\label{sigincl2}}
\end{center} 
\end{figure}

Turning now to neutrino scattering, our prediction for the CC process is 
shown in Fig.2 for scattering of  1 GeV neutrinos off carbon at $\theta=45^o$
as a function of the outgoing muon momentum $k^\prime$.
The full result obtained using the empirical scaling functions $f^{\rm QE}$ and 
$f^\Delta$ is compared with the result obtained using the RFG model $f_{RFG}$. 
It appears that the RFG cross section differs significantly from the scaling 
prediction, which lies somewhat lower and extends over a wider range in $k'$.
\begin{figure}[hbt]
\begin{center}
\vspace*{-2.5cm}
\includegraphics[scale=0.4,clip]{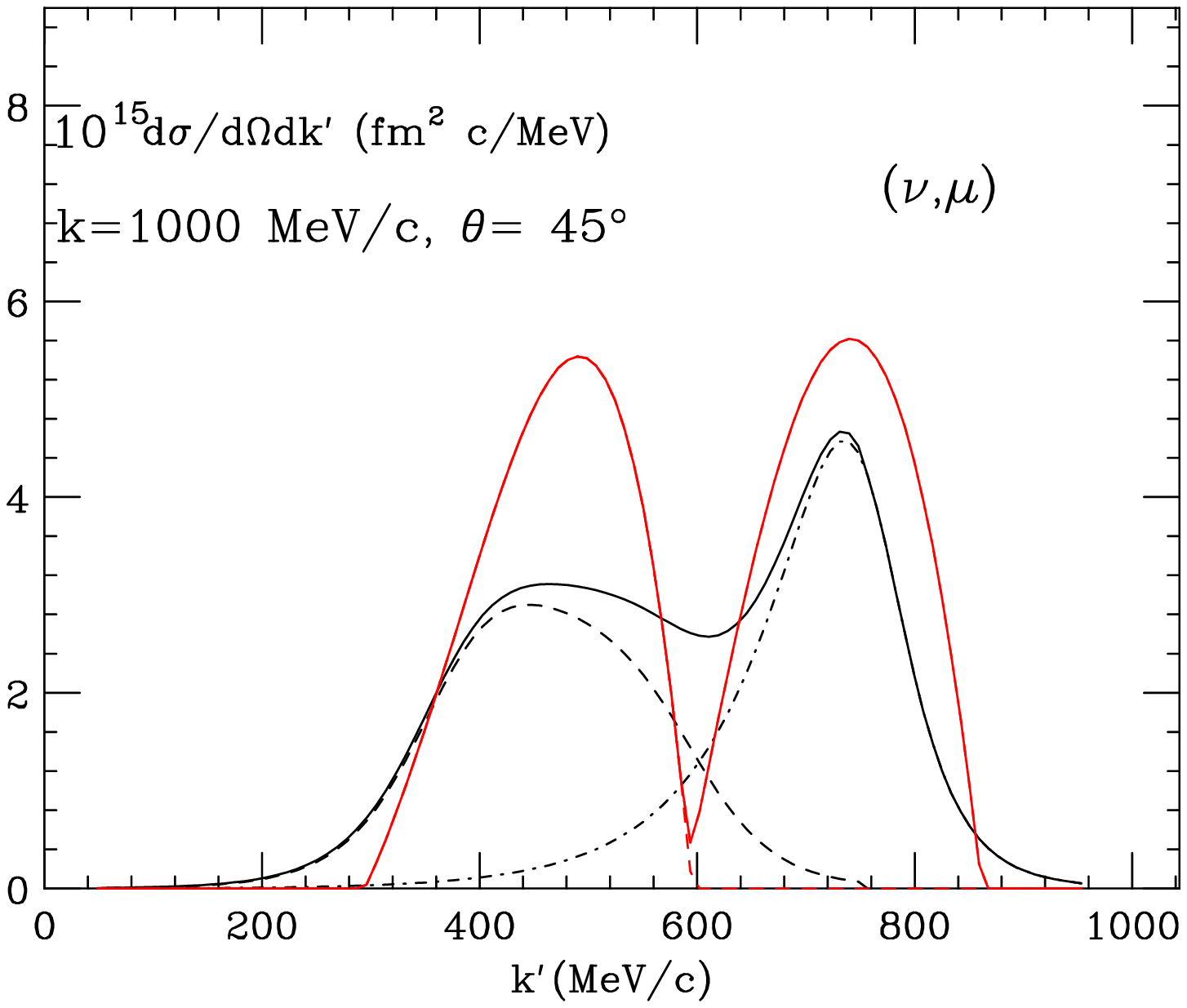}
\caption[]{Charge-changing neutrino cross section for
for 1 GeV neutrinos on $^{12}$C
and neutrino-muon scattering angles of 45 degrees. 
The cross section is plotted versus the final-state muon momentum $k'$. 
Solid line: superscaling
prediction; heavier line (red on line): RFG.
The separate  QE and $\Delta$ contribution are shown (dotted lines). 
(Ref.~\cite{Amaro:2004bs})
}
\end{center} 
\end{figure}
%
More results are shown in Ref.~\cite{Amaro:2004bs}, for different kinematics and including 
antineutrino scattering. Here we just summarize the main findings.
Concerning the relative weight of the QE and $\Delta$ contributions, we find that
at low angles the QE contribution dominates over the one coming
from the $\Delta$,  the two contributions becoming comparable for 
$\theta\sim 90^o$.
Moreover, a large difference between $\nu$ and $\bar\nu$ cross sections 
is predicted at backward angles due to the balance between the various contributions to
the total response (see Eq.~(4)). 
For both the $\nu$ and $\bar\nu$ cases the transverse and axial-transverse
responses $R_T$ and $R_{T^\prime}$ are the dominant ones, whereas the 
contributions of $R_{CC}$, $R_{CL}$ and $R_{LL}$ is less than 5\%.
However $\widehat{V}_{T} R_{T} \simeq \widehat{V}_{T^\prime} R_{T^\prime}$, so
that they interfere constructively  in $\nu$ scattering, but tend to cancel
in $\bar\nu$ scattering. As a consequence small changes in the nuclear 
model could have large effects on the antineutrino cross 
sections.

Concerning the neutral current reactions, we show in Fig.~3 (upper panel) the 
comparison
between the RFG and the superscaling results in the quasi-elastic peak. 
As in the CC case, the empirical superscaling
function predicts a lower and more extended differential cross section.

It is well-known (see, e.g., Refs.~\cite{Musolf:1993tb,Alberico:1997vh}) that
the NC reactions are very sensitive to the strangeness content
of the nucleon and are then used, together with parity-violating electron 
scattering~\cite{Musolf:1993tb,Donnelly:1991qy}, 
to measure the strange form factors of the nucleon.
We show in Fig.3 (lower panel) the cross section
obtained from the phenomenological
superscaling function in a situation where no strangeness is
assumed (solid line) with the ones obtained including strangeness
in the magnetic (long-dashed) and axial-vector (dotted) form
factors, using for $\mu_s=G_M^{(s)}(0)$ a representative value
extracted from the recent world studies of parity-violating electron
scattering and taking $g_A^s=G_A^{(s)}(0)$ to be
$-0.2$. The effects from inclusion of
electric strangeness are not shown here, since $G_E^{(s)}$ has
almost no influence on the full cross sections.

\begin{figure}[th]
\hspace{-3cm}
\includegraphics[scale=0.75, bb= 50 400 540 730]{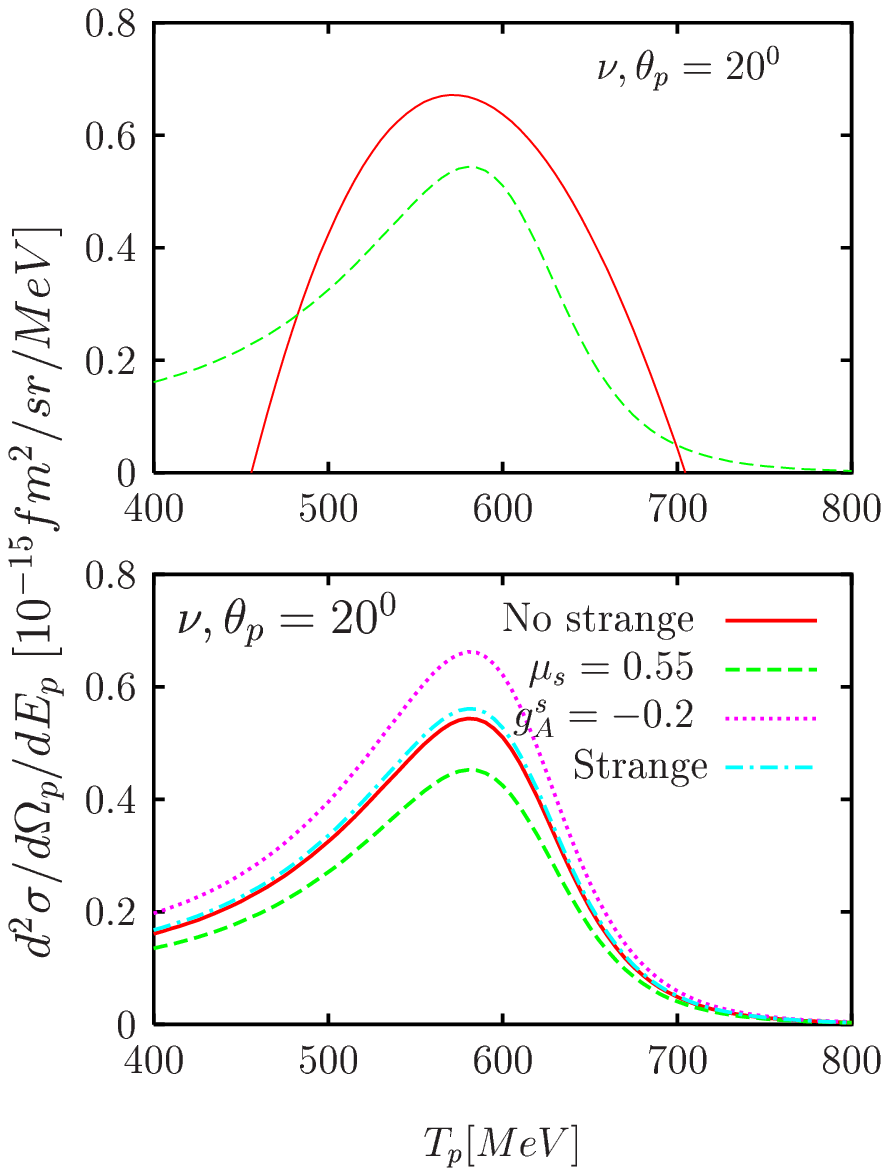}
\caption{Upper panel: quasielastic differential cross section for neutral
current neutrino scattering at 1 GeV from
$^{12}$C for proton knockout obtained using the RFG
(solid), 
and the empirical scaling function (dashed). $T_p$ is the outgoing proton
kinetic energy.
Lower panel: sensitivity to the magnetic and axial strangeness content of the nucleus.
(Ref.~\cite{NC})
\label{fig5}}
\end{figure}

\section{Conclusions}

We have illustrated a method which yields predictions for neutrino-nucleus
cross sections by extracting the scaling function from electron scattering
data in the kinematical regions of the quasi-elastic peak and of the 
$\Delta$-resonance.
The method reproduces the inclusive $(e,e')$ data in a wide range of 
kinematical conditions and for different nuclear targets. 
It has the merit of being model independent and accounts for 
that part of the nuclear dynamics which superscales. 

The cross sections obtained by using this procedure are sensibly different 
from the ones obtained in the relativistic Femi gas model, which is therefore
questionable when used in analyses of neutrino experiments.

The error implicit in this procedure is related to the violations of 
superscaling. These mainly arise from meson-exchange currents and
their associated correlations in both 
the particle-hole~\cite{Amaro:2003yd,Amaro:2001xz,Amaro:1998ta} 
and the two particle-two hole~\cite{DePace:2004cr,DePace:2003xu} sectors, 
whose impact is bound by the data to be within at most 10\%.

Of course the specific nature of the scaling function should be accounted for
by any reliable microscopic calculation. In particular the origin of
the asymmetric shape of $f^{\rm QE}$ with respect to the scaling variable is hard 
to be explained on a microscopic basis and represents a valid test of different
nuclear models~\cite{Caballero:2005sj}.

\section*{Acknowledgments}

The work presented has been carried out in collaboration with J.E.~Amaro, 
J.A.~Caballero, T.W.~Donnelly, A.~Molinari and I.~Sick.

\end{document}